\documentstyle{article}
\title{\centerline{SYMMETRIES OF PDEs SYSTEMS}
\centerline{IN SOLAR PHYSICS}
\centerline{AND CONTACT GEOMETRY}}
\author{N.B\^IL\u{A}}
\date{}
\begin{document}
\maketitle

\begin{abstract}

One considers a class of PDEs systems (1) equivalent to (8) and one determines
the PDEs system (10) which defines the associated symmetry group.
Particularly, for the Blair system (2)+(3) equivalent to (14), one finds the
symmetry Lie group $G$ (Theorem 5).
One proves that the class of PDEs systems in the form (8)+(14$'$) which are
invariant with respect to the Lie group $G$, is reduced to the Blair system
(Theorem 6). One finds new solutions $B^{(2)}_1$ and $B^{(3)}_1$
of the Blair system, and thus new "force-free" model of solar physics.

\end{abstract}

{\bf Key-words}: Blair system, infinitesimal symmetries, symmetry group,
group-invariant solutions.

{\bf Mathematical Subject Classification: 58G35, 35A30}

{\bf Physics Subject Classification: 96.60.H}

\section{Introduction}

The classical method for finding the symmetries of PDEs reduces to
the Lie group method of infinitesimal transformations. The classical symmetries
of PDEs are related to vector fields which are functions of the independent
and dependent variables. A symmetry group is a Lie group which transforms
the solutions of the system into itself, thus it provides a mean of
classifying different symmetry classes of solutions, where two solutions are
considered to be equivalent if one can be transformed into the other by
the same of the group element.
The infinitesimal generator is a vector field on the underlying
manifold which determines a flow (1-parameter group of transformations).
One can regard the entire group of symmetries
as being generated in this manner by the composition of the basic flows of
its infinitesimal generators. 
In [6] G.Bluman and J.D.Cole introduced the algorithm of finding the 
solutions of the system by using the transformations of symmetry.
A modern presentation of this theory can be found for example in [12], [15],
[19]. There are many applications of this theory in the study of the PDEs
systems which arises of mathematics, mechanics and physics, for example
[1]-[4], [6]-[19].

The solutions of the vector equations
$$
\hbox{curl}\;B=f\cdot B
\leqno(1)
$$
and
$$
\hbox{div}\;B=0,
\leqno(2)
$$
where $f$ is an arbitrary function,
and
$$
B=u(x,y,z){\partial \over {\partial x}}+v(x,y,z){\partial \over {\partial y}}
+w(x,y,z){\partial \over {\partial z}},
$$
is a differentiable vector field on a
simply connected domain $D\subset {\bf R}^3$,
define "force-free" model of solar
physics.
On the other hand, the vector equation
$$
\hbox{curl}\;B=\vert B\vert \cdot B,
\leqno(3)
$$
was introduced by Blair in [5]:
a solution of it gives a conformally flat contact metric
structure on {\bf R}$^3$.
The vector fields which satisfy the vector equation (3)
are
$$
B_1=\sin z {\partial \over {\partial x}}+
\cos z {\partial \over {\partial y}},
$$
and 
$$
B_2={8(xz-y)\over {(1+x^2+y^2+z^2)^2}}{\partial \over {\partial x}}+
{8(x+yz)\over {(1+x^2+y^2+z^2)^2}}{\partial \over {\partial y}}+
{4(1+z^2-x^2-y^2)\over {(1+x^2+y^2+z^2)^2}}{\partial \over {\partial z}},
$$
and $B_1$ is a solenoidal vector field.
In the cylindrical coordinates, by using the method of succesive
aproximations, Blair gives a solution of the equation (3) for which (2)
is also satisfied, thus a new solution for the "force-free" model equations.
The finding of the PDEs system (2)+(3) solutions is still an open problem.

In this paper we shall determine infinitesimal symmetries associated to the
PDEs system (1) for $f=f(u,v,w)$ (Theorem 3) and in the particular case for
$f=\vert B\vert $ (Theorem 4). We call {\it Blair system} the PDEs system
(2)+(3). Also we shall find the symmetry group $G$ associated to it
(Theorem 5).
We shall prove that the only PDEs system (1)+(2) which is invariant with
respect to the group $G$ is the Blair system (Theorem 6).
We shall prove that the known solutions are group-invariant solutions
and by using the symmetry group $G$ one finds new solutions of it: $B^{(2)}_1$
and $B^{(3)}_1$.

We make the remarks that there are many computational symbolic programs for
finding the defining system of infinitesimal symmetries and ours calculus was
verified by using the Head's program LIE [10].
This paper gives a new point of view for the PDEs systems which appear in the
solar physics and as well in the contact geometry.

We adopt the notations of the Olver's book [12].
We start to make a short presentation of the 
symmetry group theory in the general case of the PDEs system.

\section{Symmetry group for PDEs system}

Let us consider the PDEs system
$$
\Delta _\nu (x,u^{(n)})=0,\;\;\nu =1,...,l, \leqno(4)
$$
with $x=(x^1,...,x^p)$ the independent variables,
$u=(u^1,...,u^q)$ the dependent variables and
$$
\Delta (x,u^{(n)})=(\Delta _1(x,u^{(n)}),...,
\Delta _l(x,u^{(n)})),
$$
where we denote $ u^{(n)}$ all the partial derivatives of the function $u$
to 0 up $n$.
Any function $u=h(x)$,
$$
h:D\subset {\bf R}^p\to U\subset {\bf R}^q,\;\;
h=(h^1,...,h^q),
$$
induces the function 
$$
pr^{(n)}h:D\to U^{(n)},
$$
called {\it the n-th prolongation of $h$}, which 
for each $x\in D$, $pr^{(n)}h(x)$ is a vector whose $qp^{(n)}=qC^{n}_{p+n}$
entries represent the values of $h$ and all its derivatives up to order
$n$ at the point $x$.

The total space $D\times U^{(n)}$, whose coordinates represent the
independent variables, the dependent variables and the derivatives of the
dependent variables up to order $n$, is called {\it the n-th order
jet space} of the underlying space $D\times U$.

Thus $\Delta $ is a map from the jet space $D\times U^{(n)}$ to ${\bf R}^l$.
The PDEs system (4) determines a subvariety
$$
{\cal S}=\{(x,u^{(n)})\;\vert \Delta (x,u^{(n)})=0\}
$$
of the total jet space $D\times U^{(n)}$. One identifies the PDEs system (4)
with its corresponding subvariety ${\cal S}$.

Let $M\subset D\times U$ be an open set. We suppose $X$ is
a vector field on $M$, with corresponding local 1-parameter
group $\exp (\varepsilon X)$. {\it The n-th prolongation of X},
denoted by $pr^{(n)}X$, is a vector field on the $n$-jet space $M^{(n)}$,
and is defined to be the infinitesimal generator of the corresponding
prolonged 1-parameter group $pr^{(n)}[\exp (\varepsilon X)]$:
$$
pr^{(n)}X\vert _{(x,u^{(n)})}=
{d\over {d\varepsilon }} pr^{(n)}[\exp(\varepsilon X)](x,u^{(n)})
\vert _{\varepsilon =0}
$$
for any $(x,u^{(n)})\in M^{(n)}$.

The PDEs system (4) is called to be of {\it maximal rank} if
the Jacobian matrix
$$
J_\Delta (x,u^{(n)})=\left({\partial \Delta _\nu
\over {\partial x^i}},{\partial \Delta _{\nu }\over {\partial u^\alpha _J}}
\right)
$$
of $\Delta $, with respect to all the variables $(x,u^{(n)})$,
is of rank $l$ whenever
$$
\Delta (x,u^{(n)})=0.
$$

{\bf Theorem 1.}
{\it Let
$$
X=\sum _{i=1}^p\zeta ^i(x,u)
{\partial \over {\partial x^i}}+\sum _{\alpha =1}^q
\Phi ^{\alpha }(x,u){\partial \over {\partial u^\alpha }}
$$
be a vector field on open set $M\subset D\times U$.
The $n$-th prolongation of $X$ is the vector field
$$
pr^{(n)}X=X+\sum _{\alpha =1}^q\sum _J \Phi _{\alpha }^J(x,
u^{(n)}){\partial \over {\partial u^{\alpha }_J}}, \leqno(5)
$$
defined on the corresponding jet space $M^{(n)}\subset D\times U^{(n)}$,
the second summation being over all multi-indices
$J=(j_1,...j_k)$ with $1\leq j_k\leq p$, $1\leq k\leq n$.
The coefficient functions $\Phi ^J_{\alpha }$ of $pr^{(n)}X$ are given by the following formula
$$
\Phi ^J_{\alpha }(x,u^{(n)})=D_J\left(\phi _\alpha -
\sum _{i=1}^p\zeta ^iu^\alpha _i\right)+\sum _{i=1}^p\zeta ^iu_{J,i}^\alpha ,
$$
where $u^\alpha _i={\partial u^{\alpha } \over {\partial x^i}},\;\;
u^\alpha _{J,i}={\partial u^{\alpha }_J \over {\partial x^i}}.$}

{\bf Theorem 2}.
({\it Infinitesimal criterion of invariance}):
{\it Let us consider the PDEs system (4) of maximal rank defined over
$M\subset D\times U$.
If $G$ is a local group of transformations acting on $M$, and
$$
pr^{(n)}X[\Delta _{\nu }(x,u^{(n)})]=0,\;\;\nu =1,...,l,\leqno(6)
$$
whenever $\Delta  _\nu (x,u^{(n)})=0$, for every infinitesimal
generator
$X$ of $G$, then $G$ is a symmetry group of the system.}

{\bf Proposition 1}.
{\it If the PDEs system (4) defined on $M\subset D\times U$ is of
maximal rank,
then the set of infinitesimal symmetries of the PDEs system
forms a Lie algebra on $M$. Moreover, if this algebra is
finite-dimensional, then the symmetry group of PDEs system is a Lie
group of local transformations on $M$.}

{\bf Algorithm for determination of the symmetry group {\bf G} associated to
the PDEs system (4):}

-one considers the field $X$ on $M$ and one writes the infinitesimal
invariance condition (6);

-one eliminates any dependence between partial derivatives of the functions
$u^\alpha $, determined by the PDEs system (4);

-one writes the condition (6) like polynomials in the partial derivatives
of $u^{\alpha }$;
               
-one equates with zero the coefficients of partial derivatives of $u^{\alpha }$
in (6), written as polynomials in the derivatives of the functions $u^{\alpha }$;
it follows a PDEs system with respect to the unknown functions
$\zeta ^i,\; \phi ^\alpha $ and this system defines the Lie symmetry group $G$ of
the given PDEs system.

In general, for each $s$-parameter subgroup $H$ of the full symmetry
group $G$ associated to the PDEs system (in $p>s$ independent variables ),
there will correspond a family of group-invariant solutions.
Thus, a classification of these solutions is obtained by using an {\it optimal
system of group-invariant solutions} from which any other solution can be
derived.

{\bf Proposition 2}.
{\it If $ u=h(x)$ is an $H$-invariant solution of the PDEs
system and $g\in G$, then the composed
function $v=\tilde h(x)= g\cdot h(x)$ is a $\tilde H$-invariant
solution, where $\tilde H=gHg^{-1}$ is the conjugate subgroup under $g$.}

The problem of classifying group-invariant solutions reduces to the problem
of classifying subgroups of the full symmetry group $G$ under conjugation
and this is equivalent with the classifying subalgebras of the Lie algebra 
{\bf g} of the group $G$ under the adjoint representation.

An {\it optimal system of $s$-parameter subgroups} is a list of conjugacy
inequivalent $s$-parameter subgroups with the property that any other
subgroup is conjugate to precisely one subgroup in the list. Similarly, a
list of $s$-parameter subalgebras forms an {\it optimal system
of s-parameter subalgebras} if every $s$-parameter subalgebra of {\bf g} is
equivalent to a unique member of the list under some element of the adjoint
representation.

Thus we compute the adjoint representation $Ad\;G$ of the underlying Lie
group $G$, by using the Lie series : 
$$
Ad(exp(\varepsilon X)Y)=\sum \limits_{n=0}^{\infty} {\frac{\varepsilon ^n}{{%
n!}}}(adX)^n(Y) =Y-\varepsilon [X,Y]+{\frac{{{\varepsilon }^2}}{2}}%
[X,[X,Y]]-... \leqno(7)
$$

\section{Symmetries of PDEs systems in solar physics and contact geometry}

Let us consider
$$
B=u(x,y,z){\partial \over {\partial x}}+v(x,y,z){\partial \over {\partial y}}
+w(x,y,z){\partial \over {\partial z}},
$$
a differentiable vector field on a
simply connected domain $D\subset {\bf R}^3$.
Let 
$$
\left\{
\begin{array}{ccl}
w_y-v_z & =& uf \\
u_z-w_x & =& vf \\
v_x-u_y & =& wf,
\end{array} \right.
\leqno(8)
$$
be the PDEs system associated with the vector equation (1), where
$f$ is an arbitrary differentiable function of $u,v,w$.
We denote 
$$
\Delta _1=w_y-v_z-uf,\;\;
\Delta _2=u_z-w_x-vf,\;\;
\Delta _3=v_x-u_y-wf, 
$$
and compute the partial derivatives
$$
{\partial \Delta _1\over {\partial u}}=-f-uf_u,\;\;
{\partial \Delta _1\over {\partial v}}=-uf_v,\;\;
{\partial \Delta _1\over {\partial w}}=-uf_w,\;\;
{\partial \Delta _1\over {\partial v_z}}=-1,\;\;
{\partial \Delta _1\over {\partial w_y}}=1,
$$
$$
{\partial \Delta _2\over {\partial u}}=-vf_u,\;\;
{\partial \Delta _2\over {\partial v}}=-f-vf_v,\;\;
{\partial \Delta _2\over {\partial w}}=-vf_w,\;\;
{\partial \Delta _2\over {\partial u_z}}=1,\;\;
{\partial \Delta _2\over {\partial w_x}}=-1,
$$
$$
{\partial \Delta _3\over {\partial u}}=-wf_u,\;\;
{\partial \Delta _3\over {\partial v}}=-wf_v,\;\;
{\partial \Delta _3\over {\partial w}}=-f-wf_w,\;\;
{\partial \Delta _3\over {\partial u_y}}=-1,\;\;
{\partial \Delta _3\over {\partial v_x}}=1.
$$

Let $J_\Delta $ be the Jacobi matrix of the function $\Delta $.
It results that $rank J_{\Delta }=3$ and thus the PDEs system
(8) is of maximal rank.
Let us consider
$$
X=\zeta {\partial \over {\partial x}}+\eta {\partial \over {\partial y}}
+\theta {\partial \over {\partial z}}+
\phi {\partial \over {\partial u}}+\lambda {\partial \over {\partial v}}
+\psi {\partial \over {\partial w}},
$$
a vector field on an open set $M\subset D\times U$ of the space of the
independent and dependent variables of the system, where $\zeta ,\eta ,
\theta ,\phi ,\lambda ,\psi $ are functions of $x,y,z,u,v,w$.

If we consider $X$ to be the infinitesimal generator of the symmetry
group of PDEs system (8), then the first prolongation of it is the next
vector field
$$
pr^{(1)}X=X+\Phi ^x
{\partial \over {\partial u_x}}+\Phi ^y {\partial \over {\partial u_y}}
+\Phi ^z {\partial \over {\partial u_z}}+
\Lambda ^x{\partial \over {\partial v_x}}+\Lambda ^y{\partial \over
{\partial v_y}}+\Lambda ^z{\partial \over {\partial v_z}}+ \leqno(9)
$$
$$
+\Psi ^x{\partial \over {\partial w_x}}+\Psi ^y{\partial \over
{\partial w_y}}+\Psi ^z{\partial \over {\partial w_z}},
$$
where
$$
\begin{array}{ccl}
\Phi ^x& =& \phi _x+u_x(\phi _u-\zeta _x)-u_y\eta _x-u_z\theta _x
+\phi _vv_x+\phi _ww_x-u^2_x\zeta _u- \\
& - & u_xu_y\eta _u-u_xu_z\theta _u-u_xv_x\zeta _v-
u_yv_x\eta _v-v_xu_z\theta _v-u_xw_x\zeta _w- \\
&-& u_yw_x\eta _w-u_zw_x\theta _w,
\end{array}
$$
$$
\begin{array}{ccl}
\Phi ^y& =& \phi _y-u_x\zeta _y+u_y(\phi _u-\eta _y)-u_z\theta _y+v_y\phi _v+
w_y\phi _w-u_xu_y\zeta _u-u^2_y\eta _u- \\
\noalign{\medskip}
& -& u_yu_z\theta _u-u_xv_y\zeta _v-u_yv_y\eta _v-u_zv_y\theta _v-u_xw_y\zeta _w-u_yw_y\eta _w-
u_zw_y\theta _w,
\end{array}
$$
$$
\begin{array}{ccl}
\Phi ^z& =& \phi _z-u_x\zeta _z-u_y\eta _z+u_z(\phi _u-\theta _z)+
v_z\phi _v+w_z\phi _w-u_xu_z\zeta _u-u_yu_z\eta _u- \\
\noalign{\medskip}
& - & u^2_z\theta _u-u_xv_z\zeta _v- u_yv_z\eta _v-u_zv_z\theta _v- 
u_xw_z\zeta _w-u_yw_z\eta _w-u_zw_z\theta _w,
\end{array}
$$
$$
\begin{array}{ccl}
\Lambda ^x& =& \lambda _x+u_x\lambda _u+v_x(\lambda _v-\zeta _x)-
v_y\eta _x-v_z\theta _x+w_x\lambda _w-
u_xv_x\zeta _u-u_xv_y\eta _u- \\
\noalign{\medskip}
& - & u_xv_z\theta _u-v^2_x\zeta _v-v_xv_y\eta _v-
v_xv_z\theta _v-v_xw_x\zeta _w-v_yw_x\eta _w-v_zw_x\theta _w,
\end{array}
$$
$$
\begin{array}{ccl}
\Lambda ^y& =&  \lambda _y+u_y\lambda _u-v_x\zeta _y+v_y(\lambda _v-\eta _y)-
v_z\theta _y+\lambda _ww_y-u_yv_x\zeta _u- \\
&-&u_yv_y\eta _u-u_yv_z\theta _u-v_xv_y\zeta _v-v^2_y\eta _v-
v_yv_z\theta _v-v_xw_y\zeta _w- \\
&-&v_yw_y\eta _w-v_zw_y\theta _w,
\end{array}
$$
$$
\begin{array}{ccl}
\Lambda ^z& = & \lambda _z+u_z\lambda _u-v_x\zeta _z-v_y\eta _z+
v_z(\lambda _v-\theta _z)+w_z\lambda _w-v_xu_z\zeta _u-v_yu_z\eta _u- \\
\noalign{\medskip}
& - & v_zu_z\theta _u-v_xw_z\zeta _w-v_yw_z\eta _w-v_zw_z\theta _w-
v_xv_z\zeta _v-v_yv_z\eta _v-v^2_z\theta _v,
\end{array}
$$
$$
\begin{array}{ccl}
\Psi ^x & = & \psi _x+u_x\psi _u+v_x\psi _v+w_x(\psi _w-\zeta _x)-
w_y\eta _x-w_z\theta _x-u_xw_x\zeta _u- \\
&-&u_xw_y\eta _u-u_xw_z\theta _u-v_xw_x\zeta _v-v_xw_y\eta _v-
v_xw_z\theta _v- \\
&-&w^2_x\zeta _w-w_xw_y\eta _w-w_xw_z\theta _w,
\end{array}
$$
$$
\begin{array}{ccl}
\Psi ^y & = & \psi _y+u_y\psi _u+v_y\psi _v-w_x\zeta _y+
w_y(\psi _w-\eta _y)-w_z\theta _y-w_xu_y\zeta _u- \\
&-&w_yu_y\eta _u-w_zu_y\theta _u-w_xv_y\zeta _v-w_yv_y\eta _v-w_zv_y\theta _v-
w_xw_y\zeta _w- \\
&-&w^2_y\eta _w-w_yw_z\theta _w,
\end{array}
$$
$$
\begin{array}{ccl}
\Psi ^z & = & \psi _z+u_z\psi _u+v_z\psi _v+w_z(\psi _w-\theta _z)-
w_x\zeta _z-\eta _zw_y-u_zw_x\zeta _u-u_zw_y\eta _u- \\
&-&\theta _uu_zw_z-v_zw_x\zeta _v-v_zw_y\eta _v-v_zw_z\theta _v
-w_xw_z\zeta _w-w_zw_y\eta _w-w^2_z\theta _w.
\end{array}
$$

In this case, the infinitesimal invariance condition (6) implies
$$
\left\{\begin{array}{ccl}
-(f+uf_u)\phi -uf_v\lambda -uf_w\psi +\Psi ^y-\Lambda ^z & = & 0 \\
-vf_u\phi -(f+vf_v)\lambda -vf_w\psi  +\Phi ^z -\Psi ^x & = & 0 \\
-wf_u\phi -wf_v\lambda -(f+wf_w)\psi +\Lambda ^x-\Phi ^y & = & 0.
\end{array} \right.
$$

Substituing the functions $\Phi ^y,\Phi ^z,\Lambda ^x,\Lambda ^z,\Psi ^x,
\Psi ^y$ and after eliminating any dependencies among
the derivatives of the $u,v,w$ caused by the PDEs system (8) itself, we find
the following PDEs system
which defines the symmetry group of the studied PDEs system 
$$
\psi_y-\lambda _z-\phi (uf_u+f)-u\lambda f_v-u\psi f_w-vf\lambda _u+
wf\zeta _z+uf(\psi _w-\eta _y)+vwf^2\zeta _u-u^2f^2\eta _w+
$$
$$
+u_y(\psi _u+\zeta _z+vf\zeta _u-uf\eta _u)+v_y(\psi _v+\eta _z+vf\eta _u-
uf\eta _v)+v_z(\theta _z-\lambda _v+\psi _w-\eta _y+
$$
$$
+vf\theta _u+wf\zeta _v-2uf\eta _w)-w_x(\lambda _u+\zeta _y-wf\zeta _u+
uf\zeta _w)-w_z(\lambda _w+\theta _y-wf\zeta _w+uf\theta _w)+
$$
$$
+u_yw_z(\zeta _w-\theta _u)+v_yw_z(\eta _w-\theta _v)+u_yv_z(\zeta _v-
\eta _u)+w_xv_z(\theta _u-\zeta _w)+v_yw_x(\eta _u-\zeta _v)=0.
$$
$$
$$
$$
\phi _z-\psi _x-vf_u\phi -(vf_v+f)\lambda -vf_w\psi +vf(\phi _u-\theta _z)-
wf\psi _v+uf\eta _x-v^2f^2\theta _u+uwf^2\eta _v-
$$
$$-
u_x(\zeta _z+\psi _u+vf\zeta _u-uf\eta _u)-u_y(\eta _z+\psi _v+vf\eta _u-
uf\eta _v)+v_z(\phi _v+\eta _x-vf\theta _v+wf\eta _v)+
$$
$$
+w_x(\zeta _x-\psi _w+\phi _u-\theta _z-2vf\theta _u+wf\zeta _v+uf\eta _w)+
w_z(\phi _w+\theta _x-vf\theta _w+wf\theta _v)+
$$
$$
+u_xv_z(\eta _u-\zeta _v)+u_xw_z(\theta _u-\zeta _w)+u_yw_z(\theta _v-
\eta _w)+w^2_x(\zeta _w-\theta _u)+w_xu_y(\zeta _v-
$$
$$
-\eta _u)+w_xv_z(\eta _w-\theta _v)=0.
$$
$$
$$
$$
\lambda_x-\phi _y-w\phi f_u-w\lambda f_v-(wf_w+f)\psi +vf\theta _y+
wf(\lambda _v-\zeta _x)-uf\phi _w-w^2f^2\zeta _v+uvf^2\theta _w+
$$
$$
+u_x(\lambda _u+\zeta _y+uf\zeta _w-wf\zeta _u)+u_y(\eta _y-\phi _u+\lambda _v
-\zeta _x+vf\theta _u+uf\eta _w-2wf\zeta _v)-
$$
$$-v_y(\phi _v+\eta _x-vf\theta _v+wf\eta _v)-v_z(\theta _x+\phi _w+
wf\theta _v-vf\theta _w)+w_x(\lambda _w+\theta _y-
$$
$$
-wf\zeta _w+uf\theta _w)+u^2_y(\eta _u-\zeta _v)+u_xv_y(\zeta _v-\eta _u)
+u_xv_z(\zeta _w-\theta _u)+w_xv_y(\theta _v-\eta _w)+
$$
$$
+w_xu_y(\theta _u-\zeta _w)+u_yv_z(\eta _w-\theta _v)=0.
$$

If we equate the coefficients of the remaining unconstrained
partial derivatives of $u,v,w$ to zero, we obtain
the following PDEs system which defines the symmetry group of the studied
PDEs system:
$$
\psi_y-\lambda _z-\phi (uf_u+f)-u\lambda f_v-u\psi f_w-vf\lambda _u+
wf\zeta _z+
$$
$$
+uf(\psi _w-\eta _y)+vwf^2\zeta _u-u^2f^2\eta _w=0
$$
$$
\phi _z-\psi _x-vf_u\phi -(vf_v+f)\lambda -vf_w\psi +vf(\phi _u-\theta _z)-
$$
$$
-wf\psi _v+uf\eta _x-v^2f^2\theta _u+uwf^2\eta _v=0
$$
$$
\lambda_x-\phi _y-w\phi f_u-w\lambda f_v-(wf_w+f)\psi +vf\theta _y+
wf(\lambda _v-\zeta _x)-
$$
$$
-uf\phi _w-w^2f^2\zeta _v+uvf^2\theta _w=0
$$
$$
\phi _v=-\eta _x+f(v\eta _w-w\eta _v)
$$
$$
\phi _w=-\theta _x+f(v\theta _w-w\theta _v)      \leqno(10)
$$
$$
\lambda _u=-\zeta _y+f(w\zeta _u-u\zeta _w)
$$
$$
\lambda _w=-\theta _y+f(w\theta _u-u\theta _w)
$$
$$
\psi _u=-\zeta _z+f(u\zeta _v-v\zeta _u)
$$
$$
\psi _v=-\eta _z+f(u\eta _v-v\eta _u)
$$
$$
\phi _u-\psi _w=-\zeta _x+\theta _z+f(2v\theta _u-w\zeta _v-u\eta _w)
$$
$$
\psi _w-\lambda _v=-\theta _z+\eta _y+f(2u\eta _w-v\theta _u-w\zeta _v)
$$
$$
\eta _u=\zeta _v
$$
$$
\theta _v=\eta _w
$$
$$
\zeta _w=\theta _u.
$$

It results the next theorem

{\bf Theorem 3}.
{\it The general vector field which
describes the algebra of infinitesimal symmetries associated
to PDEs system (8) is
$$
X=\zeta {\partial \over {\partial x}}+\eta {\partial \over {\partial y}}
+\theta {\partial \over {\partial z}}+
\phi {\partial \over {\partial u}}+\lambda {\partial \over {\partial v}}
+\psi {\partial \over {\partial w}},
$$
where the functions $\zeta ,\eta ,\theta ,\phi ,\lambda $ and $\psi $
satisfy the PDEs system (10).}

Now let us consider the PDEs system
$$
\left\{
\begin{array}{ccl}
w_y-v_z & =& u\sqrt{u^2+v^2+w^2} \\
u_z-w_x & =& v\sqrt{u^2+v^2+w^2} \\
v_x-u_y & =& w\sqrt{u^2+v^2+w^2},
\end{array} \right.
\leqno(11)
$$
associated to the vector equation (3).

We obtain

{\bf Theorem 4}.
{\it A Lie algebra of infinitesimal symmetries associated to the
PDEs system (11) is described by the next vector fields}
$$
X_1=-y{\partial \over {\partial x}}+x{\partial \over {\partial y}}-
v{\partial \over {\partial u}}+u{\partial \over {\partial v}},
\;\;
X_2=-z{\partial \over {\partial y}}+y{\partial \over {\partial z}}-
w{\partial \over {\partial v}}+v{\partial \over {\partial w}}, 
$$
$$
X_3=-z{\partial \over {\partial x}}+x{\partial \over {\partial z}}
-w{\partial \over {\partial u}}+u{\partial \over {\partial w}},
\;\;
X_4={\partial \over {\partial x}},\;\;
X_5={\partial \over {\partial y}},\;\;
X_6={\partial \over {\partial z}},
$$
$$
X_7=x{\partial \over {\partial x}}+y{\partial \over {\partial y}}
+z{\partial \over {\partial z}}-u{\partial \over {\partial u}}-
v{\partial \over {\partial v}}-w{\partial \over {\partial w}},
\leqno(12)
$$
$$
X_8=2xz{\partial \over {\partial x}}+
2yz{\partial \over {\partial y}}+
(z^2-x^2-y^2){\partial \over {\partial z}}+
2(xw-zu){\partial \over {\partial u}}+
$$
$$
+2(yw-zv){\partial \over {\partial v}}-
2(xu+yv+zw){\partial \over {\partial w}},
$$
$$
X_9=xy{\partial \over {\partial x}}+
{1\over {2}}(y^2-x^2-z^2){\partial \over {\partial y}}+
yz{\partial \over {\partial z}}+
+(xv-yu){\partial \over {\partial u}}-
$$
$$
-(xu+yv+zw){\partial \over {\partial v}}+
(zv-yw){\partial \over {\partial w}},
$$
$$
X_{10}={1\over {2}}(x^2-y^2-z^2){\partial \over {\partial x}}+
xy{\partial \over {\partial y}}+
xz{\partial \over {\partial z}}-
(xu+yv+zw){\partial \over {\partial u}}+
$$
$$
+(yu-xv){\partial \over {\partial v}}+
(zu-xw){\partial \over {\partial w}}.
$$

{\bf Proof}.
We substitute $f=\sqrt{u^2+v^2+w^2}$ in the PDEs system (10).
If we consider $\zeta ,\eta ,\theta ,\phi ,\lambda $ and $\psi $
polynomial functions of $x,y,z,u,v,w$,
we get the following solution:
$$
\leqno(13)\;\;\;\;\;
\begin{array}{ccl}
\zeta & =& C_7x-C_1y-C_3z+
2C_8xz+C_9xy+{1\over {2}}C_{10}(x^2-y^2-z^2)+C_4 \\
\eta & = & C_1x+C_7y-C_2z+
2C_8yz+{1\over {2}}C_{9}(y^2-x^2-z^2)+C_{10}xy+C_5    \\
\theta & =&C_3x+C_2y++C_7z+
C_8(z^2-x^2-y^2)+C_9yz+C_{10}xz+C_6    \\
\phi &=&-C_7u-C_1v-C_3w+
2C_8(xw-zu)+C_9(xv-yu)-\\
& &-C_{10}(xu+yv+zw) \\
\lambda &=&C_1u-C_7v-C_2w+
2C_8(yw-zv)-C_9(xu+yv+zw)+\\
& &+C_{10}(yu-xv)   \\
\psi &=& C_3u+C_2v-C_7w-2C_8(xu+yv+zw)+C_9(zv-yw)+\\
& & +C_{10}(zu-xw),
\end{array}
$$
with $C_i\in {\bf R},\;i=1,...,10$.
The infinitesimal generator of the associated symmetry subgroup
is
$X=\sum ^{10}_{i=1}C_iX_i,$
where the vector fields $X_i$ are given by
the relation (12).

The structure constants of the Lie algebra generated
by the vector fields (12) are given by the next table

\vspace{0,5cm}                                                          

$$
\begin{tabular}{|c|c|c|c|c|c|c|c|c|c|c|}
\hline
$\scriptstyle{[.,.]}$ & $\scriptstyle{X_1}$ & $\scriptstyle{X_2}$ &
$\scriptstyle{X_3}$ & $\scriptstyle{X_4}$ & $\scriptstyle{X_5}$ &
$\scriptstyle{X_6}$ & $\scriptstyle{X_7}$ &
$\scriptstyle{X_8}$ & $\scriptstyle{X_9}$ & $\scriptstyle{X_{10}}$ \\
\hline
$\scriptstyle{X_1}$ & $\scriptstyle{0}$ & $\scriptstyle{X_3}$ &
$\scriptstyle{-X_2}$ & $\scriptstyle{-X_5}$ & $\scriptstyle{X_4}$ 
& $\scriptstyle{0}$ & $\scriptstyle{0}$ & $\scriptstyle{0}$ &
$\scriptstyle{X_{10}}$ & $\scriptstyle{-X_9}$
\\
\hline
$\scriptstyle{X_2}$ & 
$\scriptstyle{-X_3}$ & $\scriptstyle{0}$ & $\scriptstyle{X_1}$ &
$\scriptstyle{0}$ & $\scriptstyle{-X_6}$ & $\scriptstyle{X_5}$
& $\scriptstyle{0}$ & $\scriptstyle{2X_9}$ & $\scriptstyle{-{1\over {2}}X_8}$
& $\scriptstyle{0}$ \\
\hline
$\scriptstyle{X_3}$ & $\scriptstyle{X_2}$ & $\scriptstyle{-X_1}$ &
$\scriptstyle{0}$ & $\scriptstyle{-X_6}$ &
$\scriptstyle{0}$ & $\scriptstyle{X_4}$
& $\scriptstyle{0}$ & $\scriptstyle{2X_{10}}$ & $\scriptstyle{0}$  &
$\scriptstyle{-{1\over {2}}X_8}$ \\
\hline
$\scriptstyle{X_4}$ & $\scriptstyle{X_5}$ & $\scriptstyle{0}$ &
$\scriptstyle{X_6}$ & $\scriptstyle{0}$ & $\scriptstyle{0}$ &
$\scriptstyle{0}$
& $\scriptstyle{X_4}$ & $\scriptstyle{-2X_3}$ & $\scriptstyle{-X_1}$
& $\scriptstyle{X_7}$ \\
\hline
$\scriptstyle{X_5}$ & $\scriptstyle{-X_4}$ & $\scriptstyle{X_6}$ &
$\scriptstyle{0}$ & $\scriptstyle{0}$ & $\scriptstyle{0}$ &
$\scriptstyle{0}$
& $\scriptstyle{X_5}$ & $\scriptstyle{-2X_2}$ & $\scriptstyle{X_9}$
& $\scriptstyle{X_1}$ \\
\hline
$\scriptstyle{X_6}$ & $\scriptstyle{0}$ & $\scriptstyle{-X_5}$ &
$\scriptstyle{-X_4}$ & $\scriptstyle{0}$ & $\scriptstyle{0}$ &
$\scriptstyle{0}$
& $\scriptstyle{X_6}$ & $\scriptstyle{2X_7}$ & $\scriptstyle{X_2}$
& $\scriptstyle{X_3}$ \\
\hline
$\scriptstyle{X_7}$ & $\scriptstyle{0}$ & $\scriptstyle{0}$ &
$\scriptstyle{0}$ & $\scriptstyle{-X_4}$ & $\scriptstyle{-X_5}$ &
$\scriptstyle{-X_6}$
& $\scriptstyle{0}$ & $\scriptstyle{X_8}$ & $\scriptstyle{X_9}$
& $\scriptstyle{X_{10}}$ \\
\hline
$\scriptstyle{X_8}$ & $\scriptstyle{0}$ &
$\scriptstyle{-2X_9}$ & $\scriptstyle{-2X_{10}}$ &
$\scriptstyle{2X_3}$ & $\scriptstyle{2X_2}$ & $\scriptstyle{-2X_7}$ &
$\scriptstyle{-X_8}$
& $\scriptstyle{0}$ & $\scriptstyle{0}$ &  $\scriptstyle{0}$ \\
\hline
$\scriptstyle{X_9}$ & $\scriptstyle{-X_{10}}$ & $\scriptstyle{0}$ &
$\scriptstyle{0}$ & $\scriptstyle{X_1}$ & $\scriptstyle{-X_9}$ &
$\scriptstyle{-X_2}$
& $\scriptstyle{-X_9}$ & $\scriptstyle{0}$ & $\scriptstyle{0}$
& $\scriptstyle{0}$ \\
\hline
$\scriptstyle{X_{10}}$ & $\scriptstyle{X_9}$ & $\scriptstyle{0}$ &
$\scriptstyle{{1\over {2}}X_8}$ & $\scriptstyle{-X_7}$ & $\scriptstyle{-X_1}$ &
$\scriptstyle{-X_3}$
& $\scriptstyle{-X_{10}}$ & $\scriptstyle{0}$ & $\scriptstyle{0}$
& $\scriptstyle{0}$ \\
\hline
\end{tabular}
\vspace{0,5cm}                                                          
$$

For the Blair system
$$
\left\{
\begin{array}{ccl}
w_y-v_z & =& uf \\
u_z-w_x & =& vf \\
v_x-u_y & =& wf \\
u_x+v_y+w_z & = & 0,
\end{array} \right.
\leqno(14)
$$
where
$$
u_x+v_y+w_z = 0,
\leqno(14')
$$
is equivalent to (2),
it results

{\bf Theorem 5}.
{\it The following vector fields
$$
X_1=-y{\partial \over {\partial x}}+x{\partial \over {\partial y}}-
v{\partial \over {\partial u}}+u{\partial \over {\partial v}},
$$
$$
X_2=-z{\partial \over {\partial y}}+y{\partial \over {\partial z}}-
w{\partial \over {\partial v}}+v{\partial \over {\partial w}}, \leqno(15)
$$
$$
X_3=-z{\partial \over {\partial x}}+x{\partial \over {\partial z}}
-w{\partial \over {\partial u}}+u{\partial \over {\partial w}},
$$
$$
X_4={\partial \over {\partial x}},\;\;
X_5={\partial \over {\partial y}},\;\;
X_6={\partial \over {\partial z}}.
$$
$$
X_7=x{\partial \over {\partial x}}+y{\partial \over {\partial y}}
+z{\partial \over {\partial z}}-u{\partial \over {\partial u}}-
v{\partial \over {\partial v}}-w{\partial \over {\partial w}},
$$
describe the Lie algebra {\bf g} of the
infinitesimal symmetries associated to the Blair system (14).}

{\bf Proof}.
The PDEs system (14) is of maximal rank.
By using the above algorithm, we find
that
$$
\begin{array}{ccl}
\zeta & =& C_7x-C_1y-C_3z+C_4 \\
\eta & = & C_1x+C_7y-C_2z+C_5    \\
\theta & =&C_3x+C_2y++C_7z+C_6    \\
\phi &=&-C_7u-C_1v-C_3w \\
\lambda &=&C_1u-C_7v-C_2w  \\
\psi &=& C_3u+C_2v-C_7w,          
\end{array}
\leqno(16)
$$
where $C_i\in {\bf R},\;i=1,...,7$,
is the solution of the PDEs system which defines the symmetry group
of Blair system. It results that
$X=\sum ^{7}_{i=1}C_iX_i$,
is the infinitesimal generator of it, where the vector fiels
$X_i$ are given by (15).

{\bf Remark.}
The Lie algebra {\bf g} associated to the symmetry group of the Blair system
is a subalgebra of the algebra of infinitesimal symmetries
associated to PDEs system (11). Using the Lie series, one determines the adjoint
representation $Ad G$ of the Lie symmetry group $G$:

\vspace{0,5 cm}

$$
\begin{tabular}{|c|c|c|c|}
\hline
$\scriptstyle{Ad}$ & $\scriptstyle{X_1}$ & $\scriptstyle{X_2}$ &
$\scriptstyle{X_3}$  \\
\hline
$\scriptstyle{X_1}$ & 
$\scriptstyle{X_1}$ & $\scriptstyle{X_2\cos \varepsilon -X_3\sin \varepsilon }$
& $\scriptstyle{X_3\cos \varepsilon +X_2\sin \varepsilon }$ \\
\hline
$\scriptstyle{X_2}$ & $\scriptstyle{X_1\cos \varepsilon +X_3\sin \varepsilon }$
& $\scriptstyle{X_2}$ &
$\scriptstyle{X_3\cos \varepsilon -X_1\sin \varepsilon }$\\
\hline
$\scriptstyle{X_3}$ & $\scriptstyle{X_1\cos \varepsilon -X_2\sin \varepsilon }$
& $\scriptstyle{X_2\cos \varepsilon +X_1\sin \varepsilon }$ &
$\scriptstyle{X_3}$ \\
\hline
$\scriptstyle{X_4}$ & $\scriptstyle{X_1-\varepsilon X_5}$ &
$\scriptstyle{X_2}$ & $\scriptstyle{X_3-\varepsilon X_6}$ \\
\hline
$\scriptstyle{X_5}$ & $\scriptstyle{X_1+\varepsilon X_4}$ &
$\scriptstyle{X_2-\varepsilon X_6}$ & $\scriptstyle{X_3}$\\
\hline
$\scriptstyle{X_6}$ & $\scriptstyle{X_1}$ &
$\scriptstyle{X_2+\varepsilon X_5}$ & $\scriptstyle{X_3+\varepsilon X_4}$ \\
\hline
$\scriptstyle{X_7}$ & $\scriptstyle{X_1}$ &
$\scriptstyle{X_2}$ & $\scriptstyle{X_3}$ \\
\hline
\end{tabular}
$$

\vspace{0,5 cm}

\par
$$
\centerline{\begin{tabular}{|c|c|c|c|c|}
\hline
$\scriptstyle{Ad}$ & $\scriptstyle{X_4}$ & $\scriptstyle{X_5}$ &
$\scriptstyle{X_6}$ &
$\scriptstyle{X_7}$ \\
\hline
$\scriptstyle{X_1}$ & $\scriptstyle{X_4\cos \varepsilon +X_5\sin \varepsilon }$
& $\scriptstyle{X_5\cos \varepsilon -X_4\sin \varepsilon }$ &
$\scriptstyle{X_6}$  &
$\scriptstyle{X_7}$ \\
\hline
$\scriptstyle{X_2}$ & $\scriptstyle{X_4}$ &
$\scriptstyle{X_5\cos \varepsilon +X_6\sin \varepsilon }$ &
$\scriptstyle{X_6\cos \varepsilon -X_5\sin \varepsilon }$
& $\scriptstyle{X_7}$ \\
\hline
$\scriptstyle{X_3}$ & $\scriptstyle{X_4\cos \varepsilon +X_6\sin
\varepsilon }$ & $\scriptstyle{X_5}$ &
$\scriptstyle{X_6\cos \varepsilon -X_4\sin \varepsilon }$
& $\scriptstyle{X_7}$\\
\hline
$\scriptstyle{X_4}$ & 
$\scriptstyle{X_4}$ & $\scriptstyle{X_5}$ & $\scriptstyle{X_6}$
& $\scriptstyle{X_7-\varepsilon X_4}$\\
\hline
$\scriptstyle{X_5}$ & $\scriptstyle{X_4}$ & $\scriptstyle{X_5}$ & 
$\scriptstyle{X_6}$
& $\scriptstyle{X_7-\varepsilon{X_5}}$ \\
\hline
$\scriptstyle{X_6}$ & $\scriptstyle{X_4}$ & $\scriptstyle{X_5}$ &
$\scriptstyle{X_6}$
& $\scriptstyle{X_7-\varepsilon{X_6}}$ \\
\hline
$\scriptstyle{X_7}$ & $\scriptstyle{e^{\varepsilon }X_4}$ &
$\scriptstyle{e^{\varepsilon }X_5}$ &
$\scriptstyle{e^{\varepsilon }X_6}$ & $\scriptstyle{X_7}$ \\
\hline
\end{tabular}}
$$
\vspace{0,5cm}
\par
\noindent

{\bf Remark}.
If ($u=f(x,y,z),\;v=g(x,y,z),\;w=h(x,y,z)$) is a solution
of the Blair system, then the following triples
$$
\left\{
\begin{array}{ccl}
u^{(1)}& = & af(ax +by,-bx+ay,z)- bg(ax +by,-bx+ay,z) \\
v^{(1)} & = & bf(ax+by,-bx+ay,z) + ag(ax+by,-bx+ay,z)  \\
w^{(1)} & = & h(ax+by,-bx+ay,z),\\
\end{array} \right.
$$
$$
\left\{
\begin{array}{ccl}
u^{(2)} & = & f(x,ay+bz,-by+az ) \\
v^{(2)} & = & ag(x,ay+bz,-by+az) - bh(x,ay+bz,-by+az) \\
w^{(2)} & = & bg(x,ay+bz,-by+az)+ ah(x,ay+bz,-by+az) ,
\end{array} \right.
$$
$$
\left\{
\begin{array}{ccl}
u^{(3)} & = & af(ax+bz ,y,-bx+az) - bh(ax+bz,y,-bx+az) \\
v^{(3)} & = & g(ax+bz,y,-bx +az ) \\
w^{(3)} & = & bf(ax+bz,y,-bx+az) + ah(ax+bz,y,-bx+az) , 
\end{array} \right.
$$
$$
\leqno(17)\;\;\;\;\;\;\;\;\;\;\;\;\;\;\;\;\;\;
\left\{
\begin{array}{ccl}
u^{(4)} & = & f(x-\varepsilon ,y,z) \\
v^{(4)} & = & g(x-\varepsilon ,y,z) \\
w^{(4)} & = & h(x-\varepsilon ,y,z),
\end{array} \right.
\quad 
\quad 
\left\{
\begin{array}{ccl}
u^{(5)} & = & f(x,y-\varepsilon ,z) \\
v^{(5)} & = & g(x,y-\varepsilon ,z) \\
w^{(5)} & = & h(x,y-\varepsilon ,z),
\end{array} \right.
$$
$$
\left\{
\begin{array}{ccl}
u^{(6)} & = & f(x,y,z-\varepsilon ) \\
v^{(6)} & = & g(x,y,z-\varepsilon )\\
w^{(6)} & = & h(x,y,z-\varepsilon ), 
\end{array} \right.
\quad
\quad
\left\{
\begin{array}{ccl}
u^{(7)} & = &
e^{-\varepsilon }f(e^{-\varepsilon }x,e^{-\varepsilon }y,
e^{-\varepsilon }z) \\
v^{(7)} & = & e^{-\varepsilon }g(e^{-\varepsilon }x,e^{-\varepsilon }y,
e^{-\varepsilon }z) \\
w^{(7)} & = & e^{-\varepsilon }h(e^{-\varepsilon }x,e^{-\varepsilon }y,
e^{-\varepsilon }z), \\
\end{array} \right.
$$
are also solutions of the Blair system, where
$\varepsilon \in {\bf R}$,
$\cos \varepsilon =a$ and $\sin \varepsilon =b$.

By using the adjoint representation
$AdG$ of the grup Lie $G$ we can find an optimal system
of $s$-parameter subalgebras associated to the Blair system, with $s=1,2$.
For example, it results six 2-dimensional optimal subalgebras,
which are generated by the next vector fields:
$$
X_4,X_5;\;\;\;\;\;X_5,X_6;\;\;\;\;\;X_4,X_6,
\leqno(18)
$$
and respectively
$$
X_1,X_6;\;\;\;\;\;X_2,X_5;\;\;\;\;\;X_3,X_4.
\leqno(19)
$$

1. Let us consider the symmetry subgroup for which the
infinitesimal generators are the vector fields
$X_4$ and $X_5$ and
$$
F=G(z,u,v,w)
$$
the invariant function of it. By substituting
$$
u=g(z),\;\;\;v=h(z),\;\;\;w=k(z),
$$
in the Blair system, we get the following system of differential equations:
$$
\left\{
\begin{array}{ccl}
k=0 \\
-h'=g\sqrt{g^2+h^2} \\
g'=h\sqrt{g^2+h^2}.
\end{array}
\right.
$$
By using a coordinate transformation, the solution of it is
$$
u(z)=\sin z,\;\;\;
v(z)=\cos z,\;\;\;w=0,
$$
and it results the vector field $B_1$.

2. Now we determine the grup-invariant solutions
for the symmetry subgroup with the infinitesimal
generators $X_1$ \c{s}i $X_6$.
The invariant function is
$$
F=G(\sqrt{x^2+y^2},xu+yv,xv-yu,w).
$$
If we consider
$$
xu+yv=g(\sqrt{x^2+y^2}),\;\;\;xv-yu=h(\sqrt{x^2+y^2}),\;\;\;
w=\gamma (\sqrt{x^2+y^2}),
$$
the Blair system turns in
$$
\left\{
\begin{array}{ccl}
g=0 \\
-r\gamma '=h\sqrt{\left({h\over {r}}\right)^2+\gamma ^2} \\
h'=r\gamma \sqrt{\left({h\over {r}}\right)^2+\gamma ^2},
\end{array}
\right.
$$
where $r=\sqrt{x^2+y^2}$,
or in an equivalent form
$$
\left\{
\begin{array}{ccl}
-\gamma '=\beta \sqrt{\beta ^2+\gamma ^2} \\
{1\over {r}}\beta +\beta '=\gamma \sqrt{\beta ^2+\gamma ^2},
\end{array}
\right.
$$
by using the change of function $\beta ={h\over r}$. The solution of this
system was determined by Blair using the method of succesive aproximations [5].

By circular permutation of $x,y,z$, one determines
other solutions of the Blair system.

According with the above remark, using the relations (17),
one finds new solutions of the system.
For example in the case of the vector field
$$
B_1=\sin z {\partial \over {\partial x}}+
\cos z {\partial \over {\partial y}},
$$
the following vector fields are also solutions
$$
B^{(1)}_1=\sin (z-\varepsilon ) {\partial \over {\partial x}}+
\cos (z-\varepsilon ){\partial \over {\partial y}}=B^{(6)}_1,
$$
$$
B^{(2)}_1=\sin (az-by) {\partial \over {\partial x}}+
a\cos (az-by) {\partial \over {\partial y}}+
b\cos (az-by) {\partial \over {\partial z}},
$$
$$
B^{(3)}_1=a\sin (az-bx) {\partial \over {\partial x}}+
\cos (az-bx) {\partial \over {\partial y}}+
b\cos (az-bx) {\partial \over {\partial z}},
$$
$$
B^{(7)}_1=e^{-\varepsilon }\sin (e^{-\varepsilon }z)
{\partial \over {\partial x}}+
e^{-\varepsilon }\cos (e^{-\varepsilon }z) {\partial \over {\partial y}}.
$$
Because $B^{(4)}_1=B^{(5)}_1=B_1$, we get
the next result: the solution $B_1$ is invariant with
respect to the $2$-parameter subalgebra described by $X_4$ and $X_5$.
Analogously, one find new solutions of he Blair system
by using the Blair solution.

We make the remark that the vector field
$$
B_2={8(xz-y)\over {(1+x^2+y^2+z^2)^2}}{\partial \over {\partial x}}+
{8(x+yz)\over {(1+x^2+y^2+z^2)^2}}{\partial \over {\partial y}}+
{4(1+z^2-x^2-y^2)\over {(1+x^2+y^2+z^2)^2}}{\partial \over {\partial z}},
$$
which is a solution of the vector equation (3),
can be found by using the subgroup generated by $X_1$,
because $B^{(1)}_2=B_2$.

Now we study the inverse of the Theorem  5.

{\bf Theorem 6}.
{\it
The only PDEs system (8)+(14$'$) which is invariant with respect to
the symmetry group $G$ associated to the Blair system is the Blair system.}

{\bf Proof}.
One writes the infinitesimal conditions (6)
in the case of the PDEs system (8)+(14$'$) and
one substitutes the vector fields given by the relation (15).
It results the next PDEs system
$$
\left\{
\begin{array}{ccl}
uf_u+vf_v+wf_w=f \\
uf_v-vf_u=0   \\
vf_w-wf_v=0  \\
uf_w-wf_u=0,
\end{array}
\right.
$$
with the solution $f(u,v,w)=\sqrt{u^2+v^2+w^2}$.
$\medskip
$

We consider that the study of the PDEs system (1)+(2) from the point of view
of the symmetry group theory is very interesting, because for this PDEs system
are many known solutions, but we don't have any classification of them.

$$
$$
$$
$$
\par
\centerline{University "Politehnica" of Bucharest,}
\centerline{Departament of Mathematics I,}
\centerline{Splaiul Independen\c{t}ei, no. 313,}
\centerline{77206, Bucharest, Romania.}
\centerline{e-mail:  nbila@mathem.pub.ro}

\end{document}